# Predicting relationships between solar jet variables

Leonard A Freeman

23 Hope Street, Cambridge, CB1 3NA, UK

**Abstract**

Studies of spicules and similar solar jets reveal a strong correlation between some of the kinematic variables, particularly between the initial velocity *V*, and the subsequent deceleration, *a*. It has been proposed that there is a linear relationship between these two variables and that this offers proof for a shock wave mechanism acting on the spicules, although the linear equations found are all different. It is shown here that the relationship is better described by a non-linear, square root form: $V \propto a^{1/2}$. This relationship between *V* and *a* also provides a simple physical interpretation for the results. The different linear equations are found to be simply tangents to this *(a,V)* curve. Another method used to investigate the *(a,V)* connection is to determine the correlation coefficients between the kinematic variables from their scatter plots. It is also shown how these correlations also can be predicted from the mean value of the acceleration and height and their standard deviations for the sample under consideration. The implications of these results and the possibility that spicule behaviour is partly due to magnetic fields are discussed.

## 1. Introduction

Spicules are transient, hair-like jets that exist all over the surface of the sun. They typically reach heights of 1- 20 Mm and have lifetimes from 1 - 20 minutes or more. Recent measurements such as those made from the Hinode spacecraft or the Swedish Solar Telescope provide greater spatial and temporal resolution, allowing more accurate observations to be made of the kinematics of spicules e.g. Hansteen et al 2006; De Pontieu et al 2007a, 2007b; Anan et al 2010; Zhang et al 2012; Langangen et al 2008; Pereira et al 2012; van der Voort et al 2013; Loboda & Bogachev 2017, 2019.

These authors find a good parabolic (x,t) fit to the motion of spicules and jets, where x is the height of the spicule (or more accurately the length along its path, as spicules are usually inclined to the vertical) and t is the time from its starting point.

A spicule has an impulsive start with a maximum velocity *V*, its launch velocity. It then rises up, usually along a path inclined to the vertical. The path is considered to be a local magnetic field line. Spicules closer to the sun's poles are closer to the vertical.

After the spicule's rapid launch, it decelerates at a constant rate *a*, reaches its maximum height, *s*, when its velocity is zero. It then accelerates back down at the same rate and its final velocity is the same as its start velocity. The value of acceleration is not that due to solar gravity (274ms$^{-2}$): it can be much lower or higher, but the deceleration is constant for any particular spicule.

Strong correlations are found by the previous authors in scatter plots of the initial velocity and the deceleration. Anan et al 2010, Zhang et al 2012 and Priya et al 2018 believe that there is a proportionality between the two variables while the other authors consider that the relationship is a linear one, of the type $V = p\,a + q$, where *p* and *q* are constants. But Loboda & Bogachev 2019 showed that the actual values found for these constants vary considerably.

The proposal by Freeman 2017 that correlations may occur without the necessity of an external mechanism is further developed here, by providing a more detailed mathematical analysis of the observed correlations.

## 2. Correlations due to a restricted variable

Starting with a basic three variable equation such as $y = z\,x$, we can see how restricting the value of one variable, *z*, produces apparent correlations between the other two.

Suppose that we measure y in two different experiments. In the first case z is allowed to vary over a narrow range, between 8 and 12. In the second the values are higher, between 40 and 100 and so over a larger range. Then for random values of x and y with these restrictions, we get the plots shown in Figure 1.

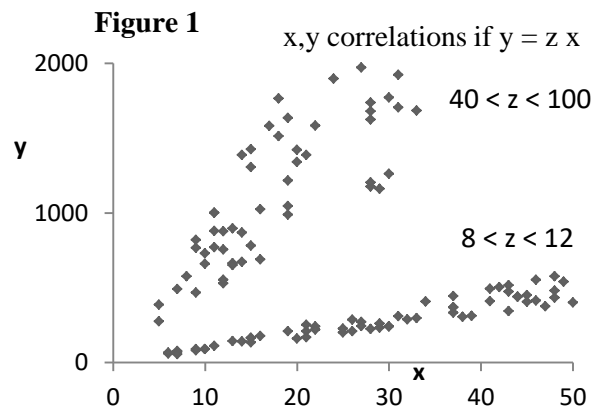

**Figure 1** x,y correlations if y = z x

It is clear that the relation between x and y depends upon the range of values of the variable z. The best fit line for the upper set of points is y = 70x, while for the lower set it is only y = 10x. The (x,y) relationship depends upon the restrictions. And the greater the spread of the points, the lower the correlation. The upper set of points in Figure 1 has a lower correlation. If there were no restrictions on z, then the whole of the graph in Figure 1 would be filled with points, and no correlation would exist.

## 3. Spicule kinematics

The simplest forms of the equations of motion for objects with constant acceleration are

$$V = at \qquad (1)$$
$$V^2 = 2as \qquad (2)$$
$$s = \tfrac{1}{2} at^2 \qquad (3)$$
$$s = \tfrac{1}{2} Vt \qquad (4)$$

where $s$ is the maximum height of a spicule. $V$ is its maximum velocity which occurs at the beginning and end of its flight. Time $t$, is the time taken to rise from the start to its maximum height. Its lifetime is then twice this value, $2t$.

These equations completely define the relationship between the four variables. We can see that the $V$-$a$ relation is only fully described when a third variable is included as in equation (1) or (2).

## 4. Experimental results

The results obtained by Loboda & Bogachev 2017 for the height and acceleration of 15 macrospicules are shown in Figure 2a with their line of best fit in Figure 2b.

Macrospicules are tall spicules, and each of the 15 that they studied had a constant deceleration. The four kinematic variables were measured for each one. The $(a,V)$ correlation can clearly be seen, even for this limited sample. The mean spicule height was 33.8 ± 6.8 Mm, and the mean deceleration 225 ± 100 ms$^{-2}$, values obtained by giving the data points equal weight.

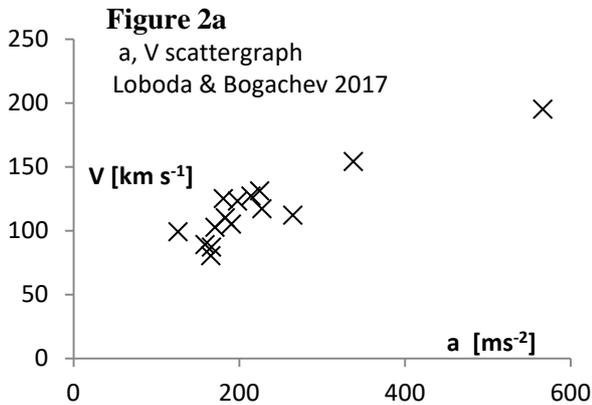

Figure 2a
a, V scattergraph
Loboda & Bogachev 2017

## 5. Scatter produced by a third variable

An $(a,V)$ graph inevitably produces scatter, because there is no single one-to-one relationship between $V$ and $a$. As can be seen from equation (2), for example, velocity is a function of two variables, only one of which is acceleration. This equation shows that $V$ depends not only on $a$, but also the height, s, which is in effect an "absent" third variable. The amount of scatter depends on the range of values for $s$. The curve of spicules of constant height, $s_0$ .can be shown on the $(a,V)$ graph. From equation (2):

$$V = \sqrt{2s_0}\sqrt{a} \qquad (5)$$

The spicules heights range from 27.0 Mm to 40.7 Mm, (two standard deviations apart). A constant height curve for each of these is displayed in Figure 2b, which shows

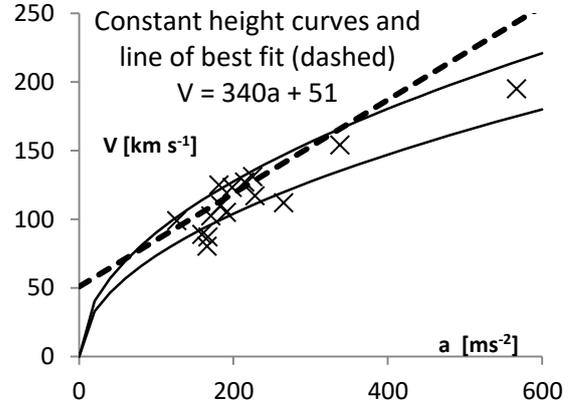

Figure 2b
Constant height curves and line of best fit (dashed)
V = 340a + 51

that most of the points lie between these two curves.

Also shown is the best fit line by Loboda & Bogachev 2017

$$V = 340a + 51 \qquad (6)$$

obtained from the scatter points. The points had different individual error bars, which is probably why the line does not appear at first sight to be the best fit.

## 6. Reproducing the equation of the best line

If we imagine in Figure (2b) that the curves are quite close together, with many data points between them, then it can be seen that the gradient of the best fit line will approach that of the tangent to the curve at the central location of all the points. The curve through this central point can be written

$$V = (2s_m)^{\tfrac{1}{2}} a^{1/2} \qquad (7)$$

Where $s_m$ is the mean height. The gradient at the central point is

$$\left(\frac{dV}{da}\right) = \left(\frac{s_m}{2a_m}\right)^{1/2} \qquad (8)$$

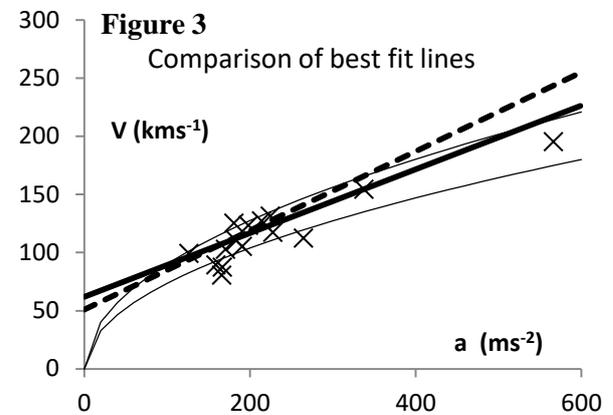

Figure 3
Comparison of best fit lines

Where $a_m$ is the mean deceleration. The line that passes through the mean point with a gradient given by equation (8), then has an intercept c, given by

$$c = \left(\frac{s_m a_m}{2}\right)^{1/2} \quad (9)$$

So the tangent equivalent for the line of best fit is

$$V = \left(\frac{s_m}{2a_m}\right)^{1/2} a + \left(\frac{s_m a_m}{2}\right)^{1/2} \quad (10)$$

For the 15 macrospicules this becomes

$$V = 274a + 62 \quad (11)$$

In reasonable agreement with equation (6). This is plotted as a solid line on Figure (3) for comparison with the line of best fit derived by Loboda & Bogachev 2017, shown dashed.

## 7. Comparison with dynamic fibrils

Dynamic fibrils are approximately 30 times smaller than macrospicules and so provide another test for the analysis produced here. De Pontieu et al 2007a recorded the motion of 257 of these. Loboda & Bogachev 2019 give the line of best fit through the 257 scatter points as

$$V = 60a + 9.7 \quad (12)$$

This can be compared to the tangent line given by equation (10). De Pontieu et al 2007a give the mean deceleration and height as 146 ms$^{-2}$ and 1250 km respectively. Substituting these values into equation (10) gives

$$V = 65a + 9.6 \quad (13)$$

This good agreement between equations (12) and (13) for these much smaller jets does confirm that the equation for the line of best fit can be found simply from the mean values for deceleration and height.

## 8. Variability of the linear equations

It can be seen how different studies have produced different linear equations for the lines of best fit on an *(a,V)* graph. The gradient of the line is given by $(s_m / 2a_m)^{1/2}$, which is simply *t/2*, so studies of spicules with long lifetimes will have higher gradients. Similarly the intercept on the *V* axis will increase when spicules with higher mean velocities are studied.

Loboda & Bogachev 2019 have found that the line of best fit varies from

$$V = 344p + 35.1 \quad (14)$$

for coronal hole jets to

$$V = 34 a + 20.9 \quad (15)$$

for type I spicules. They note that the magnetoacoustic shock model does not account for the non-zero velocity intercept.

## 9. Correlation coefficients

Correlation coefficients between pairs of variables have also been found from scatter plots. Table 1 lists these values and associated comments.

There is a good degree of consistency for some of the pairs. The *(a,V)* and *(s,V)* pairs always have a strong positive correlation, while the *(a,s)* pair has no or little connection. The others are somewhat variable, even changing from positive to negative in the case of the *(V,t)* pair.

## 10. Predicting the correlation coefficients.

It is reasonable to assume from the table that the *(a,s)* variables are unrelated. The two other variables *V* and *t* can be expressed in terms of *a* and *s*, from equations (2) and (3) to give:

$$V = (2as)^{1/2} \quad (16)$$

$$t = \left(\frac{2s}{a}\right)^{1/2} \quad (17)$$

So now *V* and *a* can be expressed as $(2as)^{1/2}$ and *a* respectively. Since these both contain the same variable *a*, we can expect a positive correlation between them, even if *a* and *s* are completely random.

The degree of correlation will depend on the range or spread of values that occur for *s* and *a*. To see this, imagine that variable *s* is almost constant. This will produce a strong positive correlation between $(2as)^{1/2}$ and *a*. But as

| Table 1 | V - a | V - t | s – V | a - t | a – s | t - s |
|---|---|---|---|---|---|---|
| Loboda & Bogachev 2019 | Strong +0.81 | +0.36 | Strong +0.67 | Strong -0.77 | Weak 0.13 corr. | Weak +0.43 |
| Pereira et al 2012 | Strong + | Strong - | | | | |
| Kiss et al 2018 | | -0.3 & 0.19 | Strong 0.8 | | | Weak, +0.2. 0.59 |
| Langangen et al | Strong + | | | | | |
| De Pontieu et al 2008 | Strong + | | | | | |
| De Pontieu et al 2007a | + linear | No clear corr. | Linear + | Weak - | No clear corr. | Linear + |
| Zhang 2012 | proportional | | + | | | |
| Kayshap 2012 | proportional + | | +0.64 | | | |
| Priya 2018 | Directly proportional +0.76 | Weak 0.04 ambiguous | | 0.02 | | |
| Gafeira | | | | | | +0.3 |

*s* becomes more variable, the *(a,V)* correlation will decrease. A similar analysis applies to the *(s,V)* variables.

Loboda & Bogachev 2019 measured the features of 330 jets mostly in coronal holes and quiet sun regions. They have produced the most comprehensive and detailed data for all four kinematic variables, giving their mean values, standard deviations, scatter plots and correlations.

The mean values and standard deviations for the decelerations and macrospicule lengths were $240.1 \pm 80.3$ ms$^{-2}$ and $25.6 \pm 5.9$ Mm respectively.

We can use these values to assign two sets of normally distributed random numbers for *a* and *s*. Then for each *(a,s)* pair *V* and *t* can be found from equations (16) and (17). This has been done for 200 *(a,s)* values and the resulting graphs of the six pairs of variables are shown in Figure 4.

The six graphs, obtained only from the mean point .($a_m = 240 \pm 80$ ms$^{-2}$, $s_m = 25.6 \pm 5.9$ Mm), closely match the scatter plots of Loboda & Bogachev 2019 and the experimental values they found for the Pearson correlation coefficients. A comparison of values is given in Table 2.

**Figure 4**

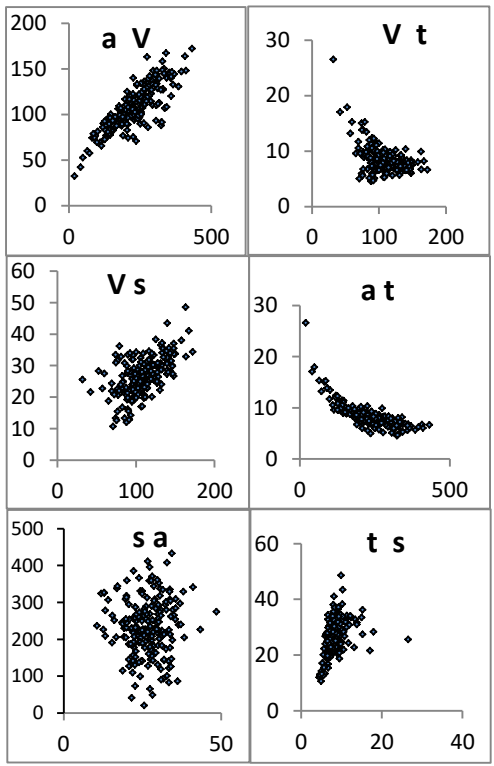

This agreement in Table 2 shows that from a single point: *($a_m, s_m$)* the approximate values for all the correlation coefficients can be found.

The units for Figure 4 are the same as for Loboda & Bogachev 2019. For the variables *(V, a, t, s)* they are (kms$^{-1}$, ms$^{-2}$, s, and Mm) respectively.

## 11.   How varying correlations arise

From Table 1 we can see that even for a single pair of variables, quite different correlations can occur: the *(V,t)* correlation has both positive and negative values. There

| Table 2 | Pearson correlation coefficients | | | | | |
|---|---|---|---|---|---|---|
| | s,a | t,s | t,a | s,V | a,V | V,t |
| A | 0.13 | 0.43 | -0.77 | 0.67 | 0.81 | -0.36 |
| B | 0.01 | 0.39 | -0.68 | 0.56 | 0.82 | -0.39 |
| A: Loboda & Bogachev 2019 | | | | | | |
| B: This work | | | | | | |

are two factors causing this, both relating to the standard deviation of the two variables. If the spread in these variables increases then generally speaking the correlation will decrease. But the ratio of the two deviations is also important. For the values shown in row B of Table 2, the relative spread of acceleration is greater than the relative spread in spicule length, resulting in a negative *(V,t)* correlation of -0.36. However, if we reverse this by doubling the standard deviation for spicule lengths then the correlation switches to positive, approximately +0.2, while leaving the other correlations approximately the same. Using the same analysis as previously shown for the *(a,V)* correlation, the relationship between *V* and *t* is really between $(sa)^{1/2}$ and $(s/a)^{1/2}$ respectively. So if the mean deviation in *a* is much less than that of *s*, we should expect a positive correlation. If the reverse applies, we get an inverse or negative correlation.

## 12.   Is the best fit linear?

Assuming that $V = k\,a^n$, where *k* is constant, then the index *n* can be found by plotting *log(V)* against *log(a)*. The gradient is *n*. For the spicule data of Loboda & Bogachev 2017 this produces the value *n = 0.54*, which is closer to the square root equation (5) than to a linear relationship. Using the same data to plot *V* against $\sqrt{a}$ gives a line with a high correlation, as shown in Figure 5. The equation of the line is approximately:

$$V = 8\sqrt{a} \qquad (18)$$

**Figure 5**

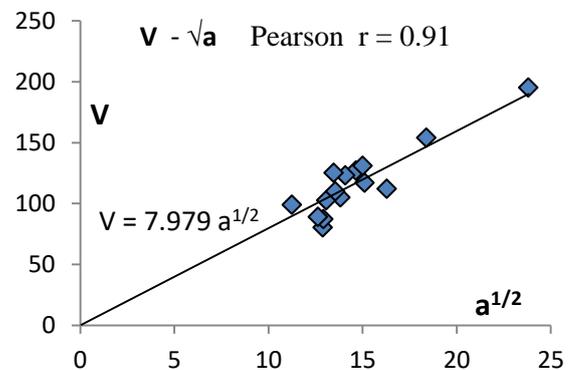

and the correlation between the variables is 0.91, which is slightly higher than for the linear equation (6) given by Loboda and Bogachev 2017. And only one constant is needed in equation (18), rather than two for a linear equation. Comparison with equation (5) gives:

$$\sqrt{(2s)} = 8 \qquad (19)$$

So the value of s is 32 Mm. This corresponds closely to the mean height of the spicules, which is 33.4 Mm. So we can conclude that the equation of best fit is just

$$V = \sqrt{(2s_m)}\sqrt{a} \qquad (20)$$

Where $s_m$ is the mean height of the spicules studied, and so we also have a physical significance for the constant.

Figure 6 shows how equation (18) matches the original *(a,V)* data points of Loboda and Bogachev 2017.

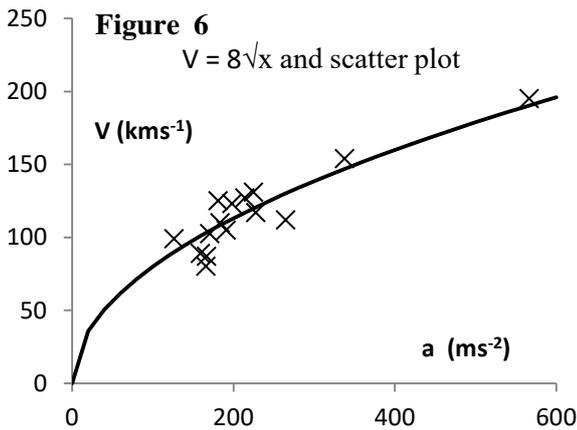

A similar analysis of the *(a,V)* results of De Pontieu et al 2007b for 37 quiet sun mottles is shown in Figure 7, where the data was taken from their figure 8. They concluded that the relationship was a linear one, but the log-log plot of the data in Figure 7 shows a gradient of n = 0.43, which again is much closer to a square root relationship than a linear one. The correlation is also better than occurs for a linear equation through the original data.

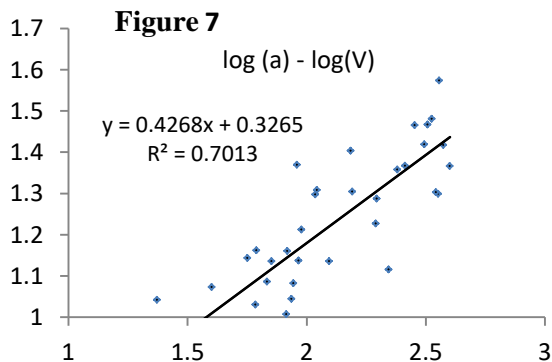

### 13. Discussion

The curve of best fit, as shown in Figure 6 may be thought of intuitively by considering a study of spicules of constant height. Then equations (2) and (5) will apply, with the deceleration of any spicule exactly proportional to $V^2$. If the condition of constant height is relaxed to allow a greater range of heights then a scatter plot similar to the points in Figure 6 will develop.

The deceleration of jets, after their initial velocity is impulsively imparted, is intriguing as it is not that due to gravity, $g_s$, at the sun's surface. For spicules it varies from less than 0.1 $g_s$ to more than 2$g_s$ . For other types of jets or filaments even higher values are found (Ji et al 2003), where filaments attracted to the solar surface may have accelerations 10 $g_s$. The authors have suggested that these very large downwards accelerations may be due to magnetic tension. If so, then it is reasonable to suppose that the same magnetic force may operate on jets such as spicules It is also generally recognised that magnetic fields control much of the behaviour of solar jets. For example their inclination traces that of the local magnetic field and the height of jets varies from one magnetic region of the sun to another. Orozco Suárez et al 2015 find that the magnetic field in spicules falls from 80G at the base to about one third this value at a height of 3000 km. Thus a typical height of the magnetic field is of the same order as spicule height, so could the local magnetic field be responsible for spicule height? Further study of magnetic field and spicule heights is needed to throw more light on this possibility.

### 14. Conclusion

It is shown here that the *(a,V)* relationship both for macrospicules and the much smaller mottles is approximately $V = \sqrt{(2s_m)}\sqrt{a}$ , where $s_m$ is the mean spicule height. This non-linearity means that the relationship does not provide the evidence that spicules are driven by magnetoacoustic shocks as is generally considered. Loboda and Bogachev 2019 have already pointed out problems with the various linear equations that occur, such as the non zero intercept on the *V* axis, suggesting that a modification to the shock mechanism may be needed. From the analysis presented here it appears that both the linear *(a,V)* relationships and the various correlations that are found are variable and arise purely from the constant acceleration equations and the statistics of the group of spicules being studied and so cannot be caused by any driving mechanism.

For constant acceleration jets there is little benefit in finding the linear equations connecting the kinematic variables, since these can all be predicted from a tangent-line to an *(a,V)* curve. It may be more useful to seek a theory which accounts for the key features of a constant sunward force acting on spicules and also for an explanation of their height.